\documentclass{article}

\usepackage[preprint]{neurips_2026}
\workshoptitle{Trustworthy AI for Good}
\PassOptionsToPackage{numbers, compress}{natbib}

\usepackage[utf8]{inputenc}
\usepackage[T1]{fontenc}
\usepackage{hyperref}
\usepackage{url}
\usepackage{booktabs}
\usepackage{amsfonts}
\usepackage{nicefrac}
\usepackage{microtype}
\usepackage{xcolor}
\usepackage{multirow}
\usepackage{graphicx}
\usepackage{tabularx}

\title{Cloned Voices, Real Consequences: \\ Evaluating Bias in Political Deepfake Detection \\ for Electoral Integrity in Brazil}

\author{%
  Lucas Rafael Gris$^{1,2,*}$,
  Daniel Casanova$^{3,*}$,
  Frederico Santos De Oliveira$^2$,
  \textbf{Alef Iury Ferreira}$^1$, \\
  \textbf{Beatriz Almeida Felício}$^1$,
  \textbf{Raul César Reis Mata}$^2$,
  \textbf{Anderson da Silva Soares}$^1$ \\[0.5em]
  $^1$Federal University of Goiás, Goiânia, Brazil \\
  $^2$Ermis, Goiânia \& São Paulo, Brazil \\
  $^3$Federal University of Technology -- Paraná, Medianeira, Brazil \\
  \texttt{\{lucas.gris, alef\_iury\_c.c, beatrizfelicio, andersonsoares\}@discente.ufg.br} \\
  \texttt{\{fred, raul\}@ermis.ai}, \texttt{danielcasanova@alunos.utfpr.edu.br} \\[0.3em]
  $^*$Equal contribution
}

\begin{document}

\maketitle

\begin{abstract}
Recent advances in generative artificial intelligence have made it easier to create convincing fake speech and spread political disinformation during elections. We introduce ParlaSpoof-BR, an audio deepfake dataset derived from recordings of the Brazilian Chamber of Deputies and extended by a range of state-of-the-art (SOTA) text-to-speech and voice conversion models. Using ParlaSpoof-BR, we benchmark SOTA audio deepfake detectors, evaluate their generalization to Brazilian Portuguese in the political domain, and examine potential biases in their predictions. Across these evaluations, our analysis shows that current systems struggle to provide consistent decisions across the diversity represented in the dataset. Differences related to the synthesis model and the extent of manipulation have a stronger effect on detector performance than demographic factors. ParlaSpoof-BR provides a domain-specific benchmark for studying audio deepfake detection in a socially consequential and underrepresented setting, supporting the development of more robust detection systems for electoral integrity in Brazil.
\end{abstract}

\section{Introduction}

Political disinformation poses a persistent threat to democratic processes by disrupting public debate, eroding trust in electoral institutions and political information, and shaping voters’ beliefs and perceptions~\cite{vaccari2020deepfakes}. Recent advances in Generative Artificial Intelligence (Generative AI) have intensified this challenge by lowering the cost of producing convincing false content at scale~\cite{pawelec2022deepfakes, twomey2023deepfake}. Audio deepfakes are particularly concerning because they can fabricate statements in the voices of political figures, allowing false narratives to shape public opinion before journalists and fact-checkers can effectively respond~\cite{pawelec2022deepfakes, twomey2023deepfake}. This risk is especially salient in Brazil, where the 2022 election was marked by coordinated disinformation campaigns~\cite{hale2024misinformation}. This risk is particularly acute in the lead-up to Brazil’s 2026 presidential election, as zero-shot voice-cloning technologies capable of generating highly realistic speech from only a few seconds of reference audio have become widely accessible~\cite{hu2026qwen3ttstechnicalreport, zhou2026voxcpm2technicalreport, omnivoice}. These developments underscore the critical importance of reliable audio deepfake detection for electoral integrity.

Yet existing audio deepfake detectors provide limited assurance in realistic settings. Although they achieve near-zero Equal Error Rates (EERs) on ASVspoof~\cite{weizman2025asvspoof}, a widely used benchmark for speech anti-spoofing, their performance degrades by an order of magnitude on real-world audio. In these conditions, detectors often rely on spurious cues, such as silence statistics, bitrate signatures, and spoken content, rather than artifacts intrinsic to speech synthesis~\cite{muller2021silence, borzi2022direction}. Detection errors also vary across gender, age, and accent~\cite{yadav2024fairssd}, while representative, Portuguese-language resources remain scarce. The only dedicated corpus, BRSpeech-DF~\cite{ferrofilho2025brspeechdf}, consists of studio-clean audiobook recordings and does not include voice conversion, partial manipulation, or regional stratification.

In this work, we introduce \textbf{ParlaSpoof-BR}\footnote{\url{https://ermisai.github.io/parlaspoof-br-demo}}, an audio deepfake benchmark based on Brazilian parliamentary speech. The dataset is publicly available for research purposes\footnote{\url{https://huggingface.co/datasets/freds0/ParlaSpoof-BR}}. It contains 2{,}000 utterances from 40 speakers balanced across gender and region, with attacks covering text-to-speech (TTS), voice conversion (VC), and semantically targeted partial manipulation. Our contributions are threefold: (1) the first Portuguese benchmark for political speech deepfakes; (2) a partial-manipulation protocol that is harder to detect than full synthesis; and (3) a systematic bias analysis showing that methodological factors have a greater impact than demographic differences.

\section{Related Work}

Audio deepfake detection has gained increasing attention as synthetic speech becomes more realistic and accessible. Prior work has examined its societal risks, detection robustness, bias, language coverage, and different forms of speech manipulation.

\textbf{Deepfakes and Elections.} 
Synthetic media pose a growing threat to electoral integrity worldwide. Pawelec~\cite{pawelec2022deepfakes} analyzes how deepfakes can undermine democratic discourse by fabricating statements attributed to candidates. Twomey \textit{et al.}~\cite{twomey2023deepfake} show that synthetic media can erode epistemic trust, meaning confidence in distinguishing reliable information from falsehoods, because their effects may occur before fact-checking can correct the record. In Brazil, the 2022 election was marked by coordinated disinformation campaigns on WhatsApp~\cite{hale2024misinformation}. More recently, manipulated and AI-generated political audio has been documented during the 2024 elections in Brazil~\cite{farrugia2024brazil}. These studies motivate domain-specific benchmarks targeting political speech rather than generic audio.

\textbf{Cross-Domain Generalization and Robustness.}
Audio deepfake detectors often degrade substantially outside their training domain. Müller \textit{et al.}~\cite{muller2026generalize} show that systems achieving strong results on ASVspoof perform considerably worse on real-world recordings, while Speech DF Arena~\cite{dowerah2026dfarena} confirms this limitation across multiple datasets and detectors. Robustness studies further show that compression, additive noise, and reverberation can suppress synthesis artifacts and increase detection errors~\cite{ali2024laundering,li2026corruption}. These findings motivate our evaluation of detectors on spontaneous Brazilian political speech under realistic acoustic perturbations.

\textbf{Bias in Deepfake Detection.} 
Detectors exhibit systematic biases that aggregate metrics conceal. FairSSD~\cite{yadav2024fairssd} audits six detectors over 0.9 million signals and finds most biased with respect to gender, age, and accent, with systematically higher false positive rates for male speakers. 

Fursule \textit{et al.}~\cite{fursule2026gender} demonstrate
statistically significant gender disparities that are obscured by
aggregate metrics. Their follow-up study~\cite{fursule2026trustworthy}
shows that group-specific thresholds can reduce false-positive-rate
disparities by 54--75\% without reducing overall detection accuracy.
Cross-lingual transfer is a persistent weakness: Liu \textit{et al.}~\cite{liu2024language} measure language mismatch across twelve languages, and Moreno \textit{et al.}~\cite{moreno2025crosslingual} show that even holding TTS architecture fixed, detection rates vary systematically by language. To the best of our knowledge, no previous audio deepfake detection study
has examined within-language regional variation as a fairness
dimension. ParlaSpoof-BR addresses this gap by considering Brazil's five geographic regions explicitly.

\textbf{Portuguese Deepfake Resources.}
Portuguese-language resources for audio deepfake detection remain limited. BRSpeech-DF~\cite{ferrofilho2025brspeechdf}, introduced as the first publicly available dataset in this setting, provides over 458{,}000 utterances covering Brazilian and European Portuguese. Its data are derived from enhanced LibriVox audiobook recordings from 62 speakers and synthetic speech generated by five zero-shot TTS systems. Evaluations on BRSpeech-DF revealed substantial cross-lingual generalization challenges, with English-trained detectors obtaining EERs between 30.67\% and 63.03\%. ParlaSpoof-BR complements this large-scale resource by targeting spontaneous Brazilian political speech and expanding the attack coverage to TTS, voice conversion, and partial manipulation, with explicit regional and gender balance.

\textbf{Partial Manipulation.} 
Fully synthesizing an utterance is unnecessary when replacing a few words suffices to invert meaning. Liu \textit{et al.}~\cite{liu2024partial} show detectors concentrate on transition regions rather than synthesis artifacts, and Gajewska \textit{et al.}~\cite{gajewska2026fraud} report 14.5\% of real fraud cases involved partial edits. Our infilling protocol extends this work by selecting targets semantically and cross-fading boundaries to suppress transition cues.

\section{Methodology}

We describe the construction of ParlaSpoof-BR, the generation of synthetic and partially manipulated speech, and the evaluation protocol used in our experiments.

\subsection{Dataset}
ParlaSpoof-BR was constructed from recordings collected between 2024-01-01 and 2026-07-01 from the official Sound Archive of the
Brazilian Chamber of Deputies\footnote{\url{https://imagem.camara.leg.br/internet/audio/}}.
These recordings preserve the reverberation and microphone variability
of parliamentary proceedings and are distributed under a Creative Commons Attribution license, allowing research and commercial use. Speaker identities
and segment boundaries were obtained directly from the archive metadata.

The dataset comprises 40 speakers, balanced by gender (20 male and
20 female), and distributed across Brazil's five geographic regions: North, Northeast, Central-West, Southeast, and South.
Eligible speakers had at least 100 segments of five seconds or longer
and 30 segments of ten seconds or longer. We selected 50 utterances per
speaker, yielding 2{,}000 bona fide samples, and five additional segments
of at least ten seconds per speaker as voice-cloning references. These
references were kept separate from the bona fide evaluation samples.
Each utterance includes the source waveform, an automatic transcription,
and word-level timestamps obtained with WhisperX~\cite{whisperx}.

\subsection{Audio Deepfake Generation}

We generate synthetic samples using pairs of distinct speakers. Speaker $A$ provides the target identity, while speaker $B$ provides the linguistic content. This speaker assignment is used consistently across the TTS and VC generation pipelines. Building on this speaker-pairing setup, we employ three complementary attack paradigms: Text-to-Speech (TTS), Voice Conversion (VC), and partial manipulation through audio infilling. The latter is evaluated both with content-preserving resynthesis and with semantically guided edits generated by a Large Language Model (LLM), which we refer to as ``LLM Attack''. Table~\ref{tab:dataset_composition} summarizes the composition of ParlaSpoof-BR across bona fide and spoofed subsets.

\textbf{Text-to-Speech.} 
We evaluate five state-of-the-art (SOTA) zero-shot voice-cloning models capable of generating audio in Brazilian Portuguese: OmniVoice~\cite{omnivoice}, XTTS-v2~\cite{casanova2024xtts}, Chatterbox Multilingual V3~\cite{resemble_chatterbox}, VoxCPM2~\cite{zhou2026voxcpm2technicalreport}, and Qwen3-TTS~\cite{hu2026qwen3ttstechnicalreport}. For each cross-speaker pair, speaker $A$ provides the target voice and speaker $B$ provides the linguistic content. Each model processes all 2{,}000 source utterances, yielding 10{,}000 TTS files.

\textbf{Voice Conversion.} 
We evaluate five zero-shot VC models: Seed-VC~\cite{seedvc}, kNN-VC~\cite{knnvc}, OpenVoice-v2~\cite{openvoice}, X-VC~\cite{zheng2026xvczeroshotstreamingvoice}, and EZVC~\cite{joglekar2025ezvceasyzeroshotanytoany}. For each source utterance, speaker $B$'s audio is converted to speaker $A$'s identity while preserving its linguistic content. Applying all five systems to the 2{,}000 source utterances yields 10{,}000 VC files.

\textbf{Audio Infilling and LLM Attack.}
For partial manipulations, we employ OmniVoice's native masked-infilling capability. Its non-causal, bidirectional architecture allows a masked region to be regenerated while conditioning on the surrounding audio and the desired transcription. This avoids generating isolated segments independently of their acoustic context.
WhisperX~\cite{whisperx} word alignments determine the corresponding
frame boundaries. 

We construct four manipulation conditions, each comprising 2{,}000 samples, yielding 8{,}000 partially manipulated audios. Three conditions correspond to fixed manipulation ratios (25\%, 50\%, and 75\% of words within the utterance), where a contiguous span of words is resynthesized while preserving the original transcription. The fourth condition employs LLM-guided semantic manipulation, where Claude Sonnet 4~\cite{anthropic2024claude} analyzes the transcript and selects a minimal, meaning-altering edit designed to change the political or factual interpretation of the utterance. The LLM chooses from five semantic categories: \textsc{antonym} (e.g., ``approved'' $\rightarrow$ ``rejected''), \textsc{number} (e.g., ``15\%'' $\rightarrow$ ``51\%''), \textsc{name} (substituting entity names), \textsc{phrase} (replacing key expressions), and \textsc{negation} (inserting or removing negation markers such as ``not''). This condition simulates targeted disinformation attacks where an adversary seeks to subtly alter the speaker's stated position.
In the three contiguous-resynthesis conditions, the original
words within spans covering 25\%, 50\%, or 75\% of the utterance are
regenerated without changing the linguistic content. The former models
realistic semantic manipulation, whereas the latter isolates acoustic
cues introduced by partial synthesis.

We additionally construct a second set of 8{,}000 partially manipulated files using the same four conditions, but restricting the manipulated regions to the first 4 seconds of each utterance. We refer to this subset as ``Short''. This constraint is motivated by the input-duration limitations of current deepfake detectors. In particular, DF-Arena processes only the first 4 seconds of an input utterance. For unconstrained partial manipulations, the altered region may therefore occur outside the portion analyzed by the detector, introducing a positional bias into the evaluation. Restricting manipulations to the initial 4 seconds ensures that the spoofed region is observable by detectors with this input constraint.

For the LLM-guided condition in the ``Short'' subset, the semantic edits are restricted to words occurring within the initial 4-second interval. Likewise, the 25\%, 50\%, and 75\%
conditions regenerate word spans selected within this interval.
This constrained variant is designed to evaluate whether detectors can
identify partial manipulations when the altered region is localized to
the beginning of the utterance. Together, the two infilling variants
yield 16{,}000 files.

For each infilling condition, the target region is masked at the token level and regenerated according to the desired transcription. The original transcription is retained for percentage-based resynthesis, while the modified transcription is used for the LLM-guided condition. Only the generated infill region is inserted back into the original waveform, preserving the unmodified portions of the recording. To reduce discontinuities at the splice boundaries, we linearly interpolate between the original and generated phase spectra over a 10 ms transition window while retaining the generated magnitude spectrum throughout the infilled region.

\subsection{Acoustic Robustness Conditions} 

We introduce three types of acoustic transformations to evaluate detector robustness under realistic processing conditions.

\textbf{Enhancement.} 
We process each of the 2{,}000 bona fide utterances with three speech
enhancement systems: Resemble Enhance\footnote{\url{https://github.com/resemble-ai/resemble-enhance}}, Demucs~\cite{defossez2020demucs}, and MetricGAN+~\cite{fu2021metricgan+}. Producing one variant per system and 6{,}000 enhanced files in total.

\textbf{Lossy Compression.}
We evaluate MP3 and OGG compression followed by decoding back to WAV. For each codec, we sample 20\% of eligible bona fide (original and enhanced) and spoofed (TTS, VC, and babble-noise) files, stratified by speaker and attack method. Partial-manipulation and voice-cloning reference files are excluded. Codec-derived files retain their original labels and identifiers. Across both codecs, this produces 35{,}200 files: 3{,}200 bona fide and 32{,}000 spoofed variants.

\textbf{Babble Noise.}
Parliamentary recordings often contain overlapping speech and background
conversations, producing interference similar to babble noise. To better
reproduce these acoustic conditions, we construct babble signals from
random speech segments extracted from other recordings using Silero
VAD~\cite{Silero_VAD}. We add this noise to the 20{,}000 full-synthesis
attacks, comprising 10{,}000 TTS and 10{,}000 VC files, at 20, 15, and
10~dB SNR. One variant is generated at each SNR, yielding 60{,}000
additional spoofed files. Babble noise is not applied to bona fide or
partially manipulated speech.

\begin{table}[!ht]
\caption{Composition of ParlaSpoof-BR.}
\label{tab:dataset_composition}
\centering
\scriptsize
\setlength{\tabcolsep}{3pt}
\renewcommand{\arraystretch}{0.95}

\begin{tabularx}{\columnwidth}{
    @{}
    l
    >{\raggedright\arraybackslash}X
    r
    @{}
}
\toprule
\textbf{Label} & \textbf{Subset} & \textbf{Files} \\
\midrule

\multirow[t]{3}{*}{\textbf{Bona fide}}
    & Original Chamber recordings
    & 2{,}000 \\
    & Enhancement variants
    & 6{,}000 \\
    & MP3/OGG to WAV variants
    & 3{,}200 \\

\addlinespace[2pt]

\multirow[t]{5}{*}{\textbf{Spoof}}
    & TTS: five systems
    & 10{,}000 \\
    & Voice conversion: five systems
    & 10{,}000 \\
    & OmniVoice speech infilling
    & 16{,}000 \\
    & Babble noise: three SNR levels
    & 60{,}000 \\
    & MP3/OGG to WAV variants
    & 32{,}000 \\

\midrule

\multicolumn{2}{@{}l}{\textbf{Bona fide total}}
    & \textbf{11{,}200} \\

\multicolumn{2}{@{}l}{\textbf{Spoof total}}
    & \textbf{128{,}000} \\

\midrule

\multicolumn{2}{@{}l}{\textbf{Total benchmark files}}
    & \textbf{139{,}200} \\

\bottomrule
\end{tabularx}
\end{table}

\subsection{Synthetic Speech Quality Assessment}
\label{sec:quality_assessment}

We assess clean synthetic speech in terms of perceptual quality,
intelligibility, and target-speaker similarity. UTMOS~\cite{utmos}
estimates perceptual quality, WER and CER are computed from
WhisperX~\cite{whisperx} transcriptions, and speaker similarity is
measured using the cosine similarity between ECAPA-TDNN
embeddings~\cite{desplanques2020ecapa}. TTS transcriptions are compared
with the synthesis text, whereas VC transcriptions are compared with
the corresponding source transcript. We exclude files modified by
babble noise or lossy codecs to avoid conflating generation quality with
acoustic degradation. Table~\ref{tab:synthesis_quality} reports medians
because the metric distributions are asymmetric. Higher UTMOS and ECAPA
scores are preferred, while lower WER and CER indicate better
intelligibility.

\begin{table}[!ht]
\caption{Synthesis quality metrics across TTS and VC generators.}
\label{tab:synthesis_quality}
\centering
% \small
\setlength{\tabcolsep}{4pt}
\renewcommand{\arraystretch}{1.0}
\begin{tabular}{llcccc}
\hline
\textbf{Type} & \textbf{Generator} &
\textbf{UTMOS$\uparrow$} &
\textbf{WER$\downarrow$} &
\textbf{CER$\downarrow$} &
\textbf{ECAPA$\uparrow$} \\
\hline

\multirow{5}{*}{TTS}
 & Chatterbox   & 2.983 & \textbf{2.6} & \textbf{0.7} & 0.802 \\
 & XTTS-v2      & 2.464 & 4.3 & 1.4 & 0.687 \\
 & OmniVoice    & 2.680 & 3.3 & 1.0 & \textbf{0.852} \\
 & Qwen3-TTS    & \textbf{3.220} & 4.0 & 1.3 & 0.847 \\
 & VoxCPM2      & 2.524 & 3.3 & 0.9 & 0.819 \\

\hline

\multirow{5}{*}{VC}
 & EZ-VC        & \textbf{2.452} & 19.6 & 10.2 & 0.753 \\
 & kNN-VC       & 1.943 & 9.7 & 3.4 & 0.762 \\
 & OpenVoice-v2 & 2.125 & 7.5 & 2.7 & 0.444 \\
 & Seed-VC      & 1.998 & 7.1 & 2.6 & \textbf{0.839} \\
 & X-VC         & 1.714 & \textbf{6.2} & \textbf{2.4} & 0.791 \\

\hline
\end{tabular}
\end{table}

Among the TTS systems, Qwen3-TTS achieves the highest UTMOS score (3.220), while Chatterbox obtains the lowest error rates (2.6\% WER and 0.7\% CER). OmniVoice achieves the highest target-speaker similarity (0.852), closely followed by Qwen3-TTS (0.847). Among the VC systems, EZ-VC obtains the highest UTMOS score (2.452), Seed-VC achieves the highest speaker similarity (0.839), and X-VC yields the lowest error rates (6.2\% WER and 2.4\% CER). Overall, the evaluated TTS systems generally achieve higher estimated perceptual quality and intelligibility than the VC systems, although no single system performs best across all evaluation dimensions.

\subsection{Detection Systems}

We evaluate three detectors representing different points on the capacity-generalization trade-off:

\textbf{AASIST}~\cite{aasist} is a graph-attention-based anti-spoofing system that operates on raw waveforms using a RawNet2-based~\cite{tak2021end} encoder. We use the standard pretrained configuration, trained on the ASVspoof 2019 Logical Access dataset, as an established baseline for measuring cross-domain generalization. 

\textbf{AASIST-L}~\cite{aasist} is the lightweight AASIST variant, containing 85{,}306 parameters while retaining the same overall spectro-temporal graph-attention design. Comparing it with the higher-capacity standard AASIST configuration examines whether model capacity within this architecture is associated with improved out-of-domain performance.

\textbf{DF-Arena-1B}~\cite{dowerah2026dfarena} is an industrial-grade transformer (1B parameters) trained on diverse multilingual deepfake corpora. It represents the best-case scenario for heterogeneous training and tests whether massive scale can overcome domain shift to Portuguese political speech.

\begin{figure*}[!hb]
\centering
\includegraphics[width=\textwidth]{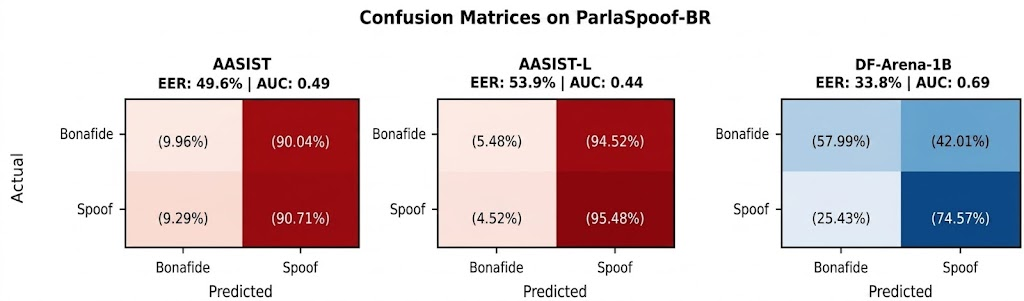}
\caption{Confusion matrices for all detectors. AASIST and AASIST-L classify nearly all samples as spoof regardless of true label. DF-Arena-1B shows better discrimination but still exhibits a 42.01\% FPR on bona fide speech}
\label{fig:confusion}
\end{figure*}

\subsection{Evaluation}

The complete benchmark comprises 139{,}200 files: 11{,}200 bona fide or bona-fide-derived files and 128{,}000 spoofed files. We divide the benchmark into a 38{,}000-file core evaluation set and a 101{,}200-file robustness set. The core set comprises 2{,}000 original bona fide recordings and 36{,}000 unperturbed attacks: 10{,}000 TTS files, 10{,}000 VC files, and 16{,}000 infilling files.

The robustness set comprises 9{,}200 bona fide-derived files and 92{,}000 spoofed files. It contains 6{,}000 enhanced bona fide recordings, 60{,}000 babble-noise spoof variants, and 35{,}200 codec-derived variants. The codec-derived conditions consist of 3{,}200 bona fide variants and 32{,}000 spoofed variants. Robustness variants are evaluated separately from the headline core-set metrics because their unequal distribution would otherwise cause the babble-noise and codec conditions to dominate the aggregate results.

For the core set, we report Equal Error Rate (EER), Area Under the ROC Curve (AUC), macro-averaged F1-score, accuracy, and spoof-class precision and recall. Robustness results are reported separately by comparing each perturbed condition with its corresponding clean condition, where applicable. All derived variants retain the identifier of their source utterance and are therefore not treated as independent source recordings.

\section{Results}

We evaluate detector performance on ParlaSpoof-BR from three complementary perspectives. We first examine overall detection performance, then analyze variation across synthesis methods and partial manipulation strategies, and finally assess how demographic and methodological factors influence detection outcomes.

\subsection{Overall Detection Performance}

Table~\ref{tab:overall_current} presents the overall performance of the three detectors on the ParlaSpoof-BR core set, which comprises 36,000 spoofed and 2,000 bona fide samples. Because spoofed samples represent 94.7\% of this set, accuracy and spoof-class precision must be interpreted together with class-balanced and threshold-independent metrics such as Macro-F1, EER, and AUC. All three detectors obtain substantially higher EERs than those reported on ASVspoof 2019 (0.83\% for AASIST, 0.99\% for AASIST-L, and 1.14\% for DF-Arena-1B), consistent with prior evidence of limited cross-domain generalization~\cite{muller2026generalize}.

\begin{table}[htbp]
\caption{Overall detection performance on ParlaSpoof-BR considering
the core subset. Prec. and Rec. denote spoof-class precision and
recall, respectively; Macro-F1 denotes the macro-averaged F1-score,
and Acc. denotes accuracy. Best results for each metric are shown in
\textbf{bold}.}
\label{tab:overall_current}
\centering
\resizebox{\columnwidth}{!}{%
\begin{tabular}{lcccccc}
\hline
\textbf{Detector}
& \textbf{EER (\%)$\downarrow$}
& \textbf{AUC$\uparrow$}
& \textbf{Prec.$\uparrow$}
& \textbf{Rec.$\uparrow$}
& \textbf{Macro-F1$\uparrow$}
& \textbf{Acc. (\%)$\uparrow$} \\
\hline
AASIST
& 49.62
& 0.493
& 0.948
& 0.907
& 0.499
& 86.45 \\

AASIST-L
& 53.95
& 0.440
& 0.948
& \textbf{0.955}
& 0.505
& \textbf{90.76} \\

DF-Arena-1B
& \textbf{33.83}
& \textbf{0.693}
& \textbf{0.970}
& 0.746
& \textbf{0.516}
& 73.73 \\
\hline
\end{tabular}%
}
\end{table}

% ORIGINAL:
% AASIST and AASIST-L perform close to chance, with EERs of 49.62\% and 53.95\% and AUCs of 0.493 and 0.440, respectively. Their high spoof recall and accuracy reflect the strong tendency of both models to predict the spoof class rather than effective class separation. As shown in Figure~\ref{fig:confusion}, this behavior results in false-positive rates of 90.04\% for AASIST and 94.52\% for AASIST-L. Accuracy is particularly misleading in this setting: spoofed samples account for 94.7\% of the core set, meaning that an always-spoof classifier would reach 94.74\% accuracy. The modest difference between the two AASIST variants further suggests that increased model capacity alone does not overcome the domain mismatch~\cite{muller2024harder}.
% DF-Arena-1B performs better, reducing the EER to 33.83\% and increasing the AUC to 0.693. It also achieves the highest spoof precision (0.970) and Macro-F1 (0.516). Its lower recall (0.746) and accuracy (73.73\%) result from predicting the spoof class less often, which reduces the advantage conferred by the class imbalance. This produces a more balanced decision pattern than the AASIST variants, although 42.01\% of bona fide samples are still misclassified. Thus, DF-Arena-1B is the strongest of the evaluated detectors on ParlaSpoof-BR, but its performance remains well below that reported on ASVspoof.

AASIST and AASIST-L perform close to chance, with EERs of 49.62\% and 53.95\% and AUCs below 0.5. As shown in Figure~\ref{fig:confusion}, both models exhibit strong bias toward predicting spoof, resulting in FPRs of 90.04\% and 94.52\%, respectively. The modest difference between variants suggests that increased capacity alone does not overcome domain mismatch~\cite{muller2024harder}. DF-Arena-1B performs better (EER 33.83\%, AUC 0.693) with more balanced predictions, although 42.01\% of bona fide samples are still misclassified. It is the strongest detector evaluated, but performance remains well below ASVspoof benchmarks.

\subsection{Performance by Synthesis Method}

Table~\ref{tab:dfarena_method_current} reports DF-Arena-1B performance by synthesis
method, including unweighted means for each attack family. On average,
the detector performs better on VC than on TTS, with a lower EER
(17.1\% versus 26.7\%), higher AUC (0.885 versus 0.777), and higher
recall (99.9\% versus 81.8\%). However, the substantial variation within
both families indicates that performance depends strongly on the
individual generator.

OpenVoice-v2 obtains the lowest EER overall (6.3\%), whereas VoxCPM2
and Qwen3-TTS produce the weakest results. Their recalls of 60.9\% and
48.2\% correspond to miss rates of 39.1\% and 51.8\%, respectively.
Although these results are consistent with previously reported cross-lingual
limitations~\cite{liu2024language,moreno2025crosslingual}, the present
experiment cannot distinguish language mismatch from other
generator-specific factors.

\begin{table}[!ht]
\caption{DF-Arena-1B performance by clean synthesis method. Mean rows report
unweighted averages across generators.}
\label{tab:dfarena_method_current}
\centering
\small
\begin{tabular}{llccc}
\hline
\textbf{Type} & \textbf{Method} &
\textbf{EER (\%)$\downarrow$} &
\textbf{AUC$\uparrow$} &
\textbf{Rec. (\%)$\uparrow$} \\
\hline
\multirow{6}{*}{VC}
 & OpenVoice-v2 & 6.3 & 0.976 & 100.0 \\
 & kNN-VC       & 16.1 & 0.902 & 100.0 \\
 & X-VC         & 16.4 & 0.896 & 100.0 \\
 & Seed-VC      & 22.6 & 0.829 & 99.8 \\
 & EZ-VC        & 24.2 & 0.821 & 99.6 \\
 & \textit{Mean} & \textit{17.1} & \textit{0.885} & \textit{99.9} \\
\hline
\multirow{6}{*}{TTS}
 & XTTS-v2      & 13.1 & 0.935 & 100.0 \\
 & Chatterbox   & 14.0 & 0.923 & 100.0 \\
 & OmniVoice    & 20.3 & 0.872 & 100.0 \\
 & VoxCPM2      & 41.1 & 0.603 & 60.9 \\
 & Qwen3-TTS    & 44.8 & 0.550 & 48.2 \\
 & \textit{Mean} & \textit{26.7} & \textit{0.777} & \textit{81.8} \\
\hline
\end{tabular}
\end{table}

\subsection{Partial Manipulation}

Table~\ref{tab:infilling_protocols_v3} reports utterance-level recall for partially manipulated speech. Under the \textit{Full} protocol, DF-Arena-1B follows its standard inference procedure and observes only the first 4 seconds of each utterance. This limits the interpretation of the results because smaller manipulations are less likely to overlap the portion of the signal available to the detector, while larger manipulations have a greater probability of appearing within this window. DF-Arena-1B recall consequently increases from 26.35\% at 25\% manipulation to 44.25\% at 50\% and 71.75\% at 75\%. The \textit{Short} subset controls for this positional effect by constraining the manipulation to the first 4 seconds while preserving the original utterance duration. Under this setting, recall increases to 61.90\% and 79.00\% for the 50\% and 75\% conditions, respectively. The LLM-guided attack remains difficult to detect, with recalls of 9.49\% on \textit{Full} and 18.41\% on \textit{Short}.

The \textit{Full Sliding} protocol increases temporal coverage by applying the detector to overlapping 4-second windows across the complete utterance and assigning the audio the maximum score obtained across all windows. DF-Arena-1B recall rises to 76.20\%, 98.00\%, and 99.75\% for the 25\%, 50\%, and 75\% conditions, respectively. This large improvement indicates that the original 4-second input restriction is an important source of error for partial manipulations. However, sliding-window inference differs from the procedure for which DF-Arena-1B was originally designed and should therefore be interpreted as an alternative evaluation strategy. Even with this broader temporal coverage, the LLM-guided attack reaches only 16.01\% recall, suggesting that its difficulty cannot be explained solely by manipulation position. One possible explanation is that these attacks modify only the short segment needed to change the meaning of the utterance, leaving most of the waveform unchanged. In addition, the infilled region is generated using the surrounding audio as context, which may reduce acoustic differences between the original and modified segments. The low recall on both \textit{Short} and \textit{Full Sliding} indicates that this difficulty cannot be explained only by the 4-second input constraint of DF-Arena-1B.

Table~\ref{tab:infilling_v3_window_accuracy} provides a complementary explanation of the AASIST results by evaluating each 4-second window independently. On the \textit{Full} subset, AASIST accuracy increases from 53.03\% at 25\% manipulation to 71.34\% at 50\% and 88.98\% at 75\%, with a similar trend for AASIST-L. This apparent improvement should not be interpreted as stronger localization of synthetic artifacts. As the manipulated proportion increases, a larger fraction of windows is labeled as fake, making the task increasingly compatible with the strong tendency of both AASIST variants to predict the spoof class. This interpretation is reinforced by the \textit{Short} subset, where manipulation is confined to the beginning of the utterance and later windows remain bona fide. In this setting, AASIST and AASIST-L achieve only 26.19\% to 39.50\% window-level accuracy. DF-Arena-1B, in contrast, remains between 78.78\% and 83.33\%, indicating substantially better discrimination between manipulated and unmanipulated regions.

\begin{table}[!ht]
\caption{Infilling detection recall (\%) on ParlaSpoof-BR. Full uses
the original detector pipeline; Full Sliding uses 4s windows with
50\% overlap and maximum-score aggregation; Short evaluates the
dedicated short-audio subset without windowing. Lower recall indicates
better evasion.}
\label{tab:infilling_protocols_v3}
\centering
% \scriptsize
\setlength{\tabcolsep}{3pt}
\renewcommand{\arraystretch}{0.9}
\begin{tabular}{llccc}
\toprule
\textbf{Protocol} & \textbf{Condition} &
\textbf{AASIST} & \textbf{AASIST-L} & \textbf{DF-Arena} \\
\midrule

\multirow{4}{*}{Full}
& 25\%       & 91.15 & 98.55 & 26.35 \\
& 50\%       & 92.00 & 98.65 & 44.25 \\
& 75\%       & 93.65 & 97.35 & 71.75 \\
& LLM Attack & 91.89 & 97.90 &  9.49 \\

\midrule

\multirow{4}{*}{Short}
& 25\%       & 93.24 & 98.05 & 25.39 \\
& 50\%       & 94.05 & 97.10 & 61.90 \\
& 75\%       & 93.85 & 96.35 & 79.00 \\
& LLM Attack & 90.70 & 96.10 & 18.41 \\

\midrule

\multirow{4}{*}{Full Sliding}
& 25\%       & 99.95 & 99.95 & 76.20 \\
& 50\%       & 99.95 &100.00 & 98.00 \\
& 75\%       &100.00 & 99.90 & 99.75 \\
& LLM Attack &100.00 & 99.95 & 16.01 \\

\bottomrule
\end{tabular}
\end{table}

\begin{table}[!ht]
\caption{Window-level accuracy (\%) on ParlaSpoof-BR. A window is labeled
as fake when it intersects any portion of the annotated infilled interval.
Predictions use a 0.5 threshold. Windows have 4s duration and 50\%
overlap.}
\label{tab:infilling_v3_window_accuracy}
\centering
% \footnotesize
\setlength{\tabcolsep}{5pt}
\renewcommand{\arraystretch}{0.95}
\begin{tabular}{llccc}
\toprule
\textbf{Subset} &
\textbf{Condition} &
\textbf{AASIST} &
\textbf{AASIST-L} &
\textbf{DF-Arena} \\
\midrule

\multirow{4}{*}{Full}
& 25\%       & 53.03 & 51.05 & \textbf{70.05} \\
& 50\%       & 71.34 & 72.69 & \textbf{73.50} \\
& 75\%       & 88.98 & \textbf{92.48} & 80.62 \\
& LLM Attack & 38.15 & 35.18 & \textbf{68.25} \\

\midrule

\multirow{4}{*}{Short}
& 25\%       & 34.56 & 26.87 & \textbf{79.03} \\
& 50\%       & 37.20 & 29.33 & \textbf{83.01} \\
& 75\%       & 39.50 & 32.33 & \textbf{83.33} \\
& LLM Attack & 33.93 & 26.19 & \textbf{78.78} \\

\bottomrule
\end{tabular}
\end{table}

\subsection{Bias Analysis}
\label{sec:bias}

We compare the variation in DF-Arena-1B performance across attack, processing, quality, and demographic factors. Table~\ref{tab:bias_summary_sliding_v3} summarizes these differences by reporting the performance gap within each factor. For methodological factors, the gap is computed as the difference between the highest and lowest recall observed across the corresponding conditions. For demographic factors, it is computed as the difference between the highest and lowest EER. All gaps are reported in percentage points.

\begin{table}[htbp]
\caption{Bias factors ranked by impact on DF-Arena-1B detection using
4s sliding windows with 50\% overlap. Partial manipulation was
recalculated using the Full subset; the remaining factors use the
unchanged sliding-window benchmark results.}
\label{tab:bias_summary_sliding_v3}
\centering
\small
\begin{tabular}{llr}
\hline
\textbf{Rank} & \textbf{Bias Factor} & \textbf{Gap (pp)} \\
\hline
1 & Partial Manipulation (infilling) & 83.7 \\
2 & Synthesis Model & 51.8 \\
3 & Processing Group & 32.9 \\
4 & UTMOS Quality & 29.5 \\
5 & Speaker Similarity (ECAPA) & 19.0 \\
6 & Codec & 10.4 \\
7 & SNR Level & 9.4 \\
8 & Gender & 2.9 \\
9 & Region & 2.3 \\
\hline
\end{tabular}
\end{table}

The largest differences are associated with how fake audio is generated: partial manipulation shows the largest gap (83.7 pp), followed by synthesis model (51.8 pp). DF-Arena-1B recall ranges from 48.2\% for Qwen3-TTS to 100.0\% for OpenVoice-v2, showing that performance depends strongly on the generation method.

Audio processing also affects detection. At 10~dB SNR, 22.2\% of previously detected deepfakes escape detection. Resemble Enhance causes 98.9\% of bona fide recordings to be misclassified as fake. Compression has a strong effect, with false positive rates of 18.7\% after MP3 and 94.8\% after OGG conversion.

Demographic variation is considerably smaller: gender produces a gap of 2.9 pp and region 2.3 pp. Overall, the detector is more sensitive to how audio is generated or processed than to speaker demographics.

\section{Conclusion}

We introduced ParlaSpoof-BR, an audio deepfake benchmark comprising 139{,}200 files from 40 speakers balanced across gender and geographic region. Our evaluation shows that current detectors generalize poorly to Brazilian parliamentary speech and that performance varies substantially across synthesis methods, manipulation strategies, and audio processing conditions. Partial manipulation is particularly challenging, with the LLM-guided attack reaching only 9.49\% recall. The bias analysis shows that methodological factors produce much larger performance differences than demographics (83.7 pp for partial manipulation and 51.8 pp for synthesis model, compared with 2.9 pp for gender and 2.3 pp for region). These findings highlight the need to evaluate detectors under diverse generation and processing conditions. Future work should prioritize bias-aware training, adaptation to Brazilian Portuguese political speech, and detection methods effective on partially manipulated utterances. ParlaSpoof-BR supports research toward more reliable deepfake detection for electoral integrity in Brazil.

\section*{Acknowledgments} 

This work has been funded by the project Research and Development of Genese Digital: Scaling Interactive and Culturally Adapted Digital Humans with Generative AI, supported by the Advanced Knowledge Center in Immersive Technologies (AKCIT), with financial resources from the PPI IoT/Manufatura 4.0 / PPI HardwareBR of the MCTI, grant number 057/2023, signed with EMBRAPII. The authors also acknowledge the support and contributions of Huglabs and Ermis.ai.

\bibliographystyle{plainnat}
\bibliography{refs}

\end{document}